\documentclass{kapproc} 
\usepackage{epsfig,epsf}

\kluwerbib

\newcommand{\be}{\begin{equation}}
\newcommand{\ee}{\end{equation}}
\newcommand{\bea}{\begin{eqnarray}}
\newcommand{\eea}{\end{eqnarray}}
\newskip\humongous \humongous=0pt plus 1000pt minus 1000pt
\def\caja{\mathsurround=0pt}
\def\eqalign#1{\,\vcenter{\openup1\jot \caja
    \ialign{\strut \hfil$\displaystyle{##}$&$
    \displaystyle{{}##}$\hfil\crcr#1\crcr}}\,}
\newif\ifdtup

\normallatexbib 
\begin{document}

\articletitle{Scale Dependence of Dark Energy Antigravity} 
\author{L. Perivolaropoulos\footnote{Contributed talk at the 2nd Hellenic 
Cosmology Workshop at NOA (Athens) Jan. 2001. Based on work done 
in collaboration with M. Axenides and E. Floratos\cite{afp00}}} 
\affil{Institute of Nuclear Physics, \\ National Centre for 
Scientific Research ``Demokritos N.C.S.R.'',\\ Athens, Greece} 
\email{leandros@mail.demokritos.gr} 


\begin{abstract}
We investigate the effects of negative pressure induced by dark 
energy (cosmological constant or quintessence) on the dynamics at 
various astrophysical scales. Negative pressure induces a 
repulsive term (antigravity) in Newton's law which dominates on 
large scales. Assuming a value of the cosmological constant 
consistent with the recent SnIa data we determine the critical 
scale $r_c$ beyond which antigravity dominates the dynamics ($r_c 
\sim 1Mpc $) and discuss some of the dynamical effects implied. We 
show that dynamically induced mass estimates on the scale of the 
Local Group and beyond are significantly modified due to negative 
pressure. We also briefly discuss possible dynamical tests (eg 
effects on local Hubble flow) that can be applied on relatively 
small scales (a few $Mpc$) to determine the density and equation 
of state of dark energy.   
\end{abstract}

\begin{keywords}
Cosmological Constant, Newton's law, Dark Energy, Galactic 
Dynamics, Hubble Flow 
\end{keywords}

\section*{Introduction}
Recent cosmological observations extending the Hubble diagram of 
high redshift Type Ia supernonovae performed independetly by  two 
groups (the Supernova Cosmology Project \cite{scpc99} and the 
High-Z Supernova Team \cite{hz98,Riess:1998cb} presented evidence 
that the expansion of the universe is accelerating rather than 
slowing down. This fact combined with the Boomerang and Maxima-1 
measurements of the first acoustic peak location in the angular 
power spectrum of the cosmic microwave background (CMB) 
\cite{db00,j00} point towards a standard cosmological model with 
critical density ($\Omega = \Omega_m + \Omega_{\Lambda} = 1$) and 
a dominant $\Lambda$-like, ``dark energy" component at the present 
epoch ($\Omega_{\Lambda} \approx 0.7$). This component could be 
produced by  non-zero and positive cosmological constant $\Lambda$ 
with \be \Lambda \simeq 10^{-52} m^{-2} \label{valcc} \ee Such a 
term can produce the required repulsive force to explain the 
accelerating universe phenomenon. A diverse set of other 
cosmological observations also compellingly suggest that the 
universe posesses a nonzero negative pressure component 
corresponding to vacuum energy density of the same order as the 
matter energy 
density\cite{Weinberg:1996xe,Krauss:1995yb,Carroll:1992mt}.  

In addition to causing an acceleration to the expansion of the 
universe the existence of a non-zero cosmological constant would 
have interesting gravitational effects on various astrophysical 
scales\cite{Ostriker:1995su,afp00}. For example it would affect 
gravitational lensing statistics of extragalactic 
surveys\cite{Quast:1999fh}, large scale velocity 
flows\cite{Zehavi:1999fm} and there have been some claims that 
even smaller systems (galactic\cite{Whitehouse:1999rs} and 
planetary\cite{ct98}) could be affected in an observable way by 
the presence of a cosmological constant consistent with 
cosmological expectations. Even though some of these claims were 
falsified\cite{Wright:1998bc,Neupane:1999hr,Roberts:1987ch} the 
scale dependence of the dynamical effects of vacuum energy remains 
an interesting open issue. 
 
The effects of the vacuum energy on cosmological scales and on 
local dynamics can be obtained from the Einstein equations which 
in the presence of a non-zero cosmological constant are written as 
\be
R_{\mu \nu}-{1\over 2} g_{\mu \nu} R = 8 \pi G T_{\mu \nu} 
\label{einst}\ee 
 

These equations, under the assumptions of spherical symmetric
energy-momentum (EM) tensor $T_k^i = \rho_Q c^2\; 
diag(1,-w,-w,-w)$ and a mixture of dust-like matter 
($\rho=\rho_m$, $w=0$) and dark energy ($\rho=\rho_Q$, $-{1\over 
3} \leq w \leq -1$) lead (for the 1-1 component) to the 
generalized Newton's equation 
\begin{equation}
\ddot{r} = - G M_{eff}/r^2; \;\;\; M_{eff} = M_m (r) + M_Q(r)\;. 
\label{accel} 
\end{equation}
where $ M(r) = \frac{4\pi}{3} (1+3w)\rho r^3$. Notice that for 
$w<-{1\over 3}$ we have negative gravitating effective mass 
(antigravity) which can lead to accelerated cosmological expansion 
and to non-trivial dynamical effects on astrophysical scales. The 
accelerated cosmological expansion is obtained for $w<-{1\over 3}$ 
from the Friedman equations which for $k=0$ imply \bea \rho_Q 
&\sim & R_Q^{-3(1+w)} \\  R_Q & \sim & t^{{2\over {3 (1+w)}}} \eea 
where $R_Q$ is the scale factor of the universe. In what follows 
we focus on the effects of dark energy with $w=-1$ (cosmological 
cosntant). The more general case of $-1 < w \leq -{1\over 3}$ 
(quintessence\cite{quint}) will be discussed elsewhere. 
 
The vacuum energy implied from eq. (\ref{valcc}) ($10^{-10} 
erg/cm^3$) is less by many orders of magnitude than any sensible 
estimate based on particle physics. In addition, the matter 
density $\rho_m$  and and the vacuum energy $\rho_\Lambda$  evolve 
at different rates, with $\rho_m / \rho_\Lambda \simeq R^{-3}$ and 
it would seem quite unlikely that they would differ today by a 
factor of order unity. Interesting attempts have been made during 
the past few years to justify this apparent fine tuning by 
incorporating evolving scalar fields (quintessence\cite{quint}) or 
probabilistic arguments based on the anthropic 
principle\cite{Garriga:1999bf}). 

  
For $w=-1$ we have
 Einstein's cosmological constant with
$\Lambda = 8 \pi G \rho_{\Lambda}/c^2$ (cosmological vacuum with 
$w=-1$) and  the gravitating mass is  $M_{\Lambda} = -8\pi 
\rho_{\Lambda}r^3 /3$. Thus the generalized Newtonian potential 
leads to a gravitational interaction acceleration 


\be {\ddot r}=-{{GM}\over r^2} + {{\Lambda c^2}\over 3} r 
 \label{acc}\ee This generalized 
force includes a repulsive term \be g_r = {{\Lambda c^2}\over 3} r 
 \label{rep}\ee which is expected to dominate at distances larger than \be r_c 
= ({{3 G M}\over {\Lambda c^2}})^{1\over 3}\simeq 10^2 ({{{\bar 
M}_1}\over {{\bar \Lambda_{52}}}})^{1\over 3} pc \simeq 2\times 
10^7 ({{\bar M}_1\over {{\bar \Lambda_{52}}}})^{1\over 3} AU 
 \label{crdist}\ee 
where ${\bar M}_1$ is the mass within a sphere of radius $r_c$  in 
units of solar masses $M_\odot = 2 \times 10^{30} kg$ and  ${\bar 
\Lambda_{52}}$ is the cosmological constant in units of $10^{-52} 
m^{-2}$.  

The question we address in this report is the following: `What are 
the effects of the additional repulsive force $g_r$ on the various 
astrophysical scales?' This issue has been addressed in the 
literature for particular scales. For example it was 
shown\cite{Wright:1998bc} that the effects of this term in the 
solar system could only become measurable (by modifying the 
perihelia precession) if the cosmological constant were fourteen 
orders of magnitude larger than the value implied by the SnIa 
observations. 

In the next section it will be shown that the vacuum energy 
required to close the universe (eq. (\ref{valcc})) has negligible 
effects on the dynamics of galactic scales (few tens of kpc). The 
dynamically derived mass to light ratios of galaxies obtained from 
velocity measurements on galactic scales are modified by less than 
$0.1\%$ due to the vacuum energy term of eq. (\ref{valcc}). This 
is not true however on cluster scales or larger. Even on the 
scales of the Local Group of galaxies (about 1Mpc) the 
gravitational effects of the vacuum energy are significant. We 
show that the dynamically obtained masses of M31 and the Milky Way 
must be increased by about 35\% to compensate the repulsion of the 
vacuum energy of eq. (\ref{rep}) and produce the observed relative 
velocity of the members of the Local Group. The effects of vacuum 
energy are even more important on larger scales (rich cluster and 
supercluster). 

\section{Scale Dependence of Antigravity}
In order to obtain a feeling of the relative importance of 
antigravity vs gravity on the various astrophysical scales it is 
convenient to consider the ratio of the corresponding two terms in 
eq. (\ref{acc}). This ratio $q$ may be written as \be q={{\Lambda 
c^2 r^3}\over {3 G M}} \simeq 0.5 \times 10^{-5} {{{\bar 
\Lambda}_{52}{\bar r}_1^3}\over {{\bar M}_1}}  \label{qrat}\ee  
where ${\bar r}_1$ is the distance measured in units of $pc$. For 
the solar system (${\bar r}_1\simeq 10^{-5}$, ${\bar M}_1 = 1$) we 
have $q_{ss}\simeq 10^{-20}$ which justifies the fact that 
interplanetary measures can not give any useful bound on the 
cosmological constant. 

For a galactic system (${\bar r}_1\simeq 10^{4}$, ${\bar M}_1 = 
10^{10}$) we have $q_g \simeq 5 \times 10^{-4}$ which indicates 
that up to galactic scales the dynamical effects of the 
antigravity induced by $\Lambda$ are negligible. On a cluster 
however (${\bar r}_1\simeq 10^{7}$, ${\bar M}_1 = 10^{14}$) we 
obtain $q_c \simeq O(1)$ and the gravitational effects of the 
vacuum energy become significant. This will be demonstrated in a 
more quantitative way in what follows.  

The precessions of the perihelia of the planets provide one of the 
most sensitive Solar System tests for the cosmological constant. 
The additional precession due to the cosmological constant can be 
shown\cite{Wright:1998bc} to be \be \Delta \phi_\Lambda = 6\pi q 
\; rad/orbit  \label{addprec}\ee where $q$ is given by eq. 
(\ref{qrat}).  For Mercury we have ${\bar r}_1 \simeq 10^{-6}$ 
which leads to $q_{mc} \simeq 10^{-23}$ and $\Delta \phi_\Lambda 
\simeq 10^{-22} rad/orbit $. The uncertainty in the observed 
precession of the perihelion of Mercury is $0.1''$ per century or 
$\Delta \phi_{unc} \simeq  10^{-9} rad/orbit $ which is $13$ 
orders of magnitude larger than the one required for the detection 
of a cosmologicaly  interesting value for the cosmological 
constant. The precession per century\footnote{The angular velocity 
is smaller for distant planets and therefore the precession per 
century does not scale like the ${\bar r}_1^3$ as the precession 
per orbit does} scales like ${\bar r}_1^{3/2}$ and therefore the 
predicted additional precession per century for distant planets 
(${\bar r}_1 (Pluto) \simeq 10^2 {\bar r}_1 (Mercury)$)  due to 
the cosmological constant increases by up to 3 orders of 
magnitude.  It remains however approximatelly 10 orders of 
magnitude smaller than the precession required to give a 
cosmologically interesting detection of the cosmological constant 
even with the best quality of presently available observations.  
It is therefore clear that since the relative importance of the 
gravitational contribution is inversely proportional to the mean 
matter density on the scale considered, a cosmological constant 
could only have detectable gravitational effects on scales much 
larger than the scale of the solar system. 

On galactic scales, the rotation velocities of spiral galaxies as 
measured in the 21cm line of neutral hydrogen comprise a good set 
of data for identifying the role of the vacuum energy. This is 
because these velocity fields usually extend well beyond the 
optical image of the galaxy on scales where the effects of 
$\Lambda$ are maximized and because gas on very nearly circular  
orbits is a precise probe of the radial force law. For a stable 
circular orbit with velocity $v_c$ at a distance $r$ from the 
center of a galaxy with mass $M$ we obtain using eq. (\ref{acc}) 
\be v_c^2 = {{GM}\over r} -{{\Lambda c^2 r^2}\over 3}  
\label{rotvel} \ee We now define the rescaled dimensionless 
quantities  ${\bar v}_{100}$, ${\bar r}_{10}$ and ${\bar M}_{10}$ 
as follows:  
\begin{equation}
\eqalign{ 
 v_c &= 100 \; {\bar v}_{100} \; km/sec \cr
 r &= 10 \;{\bar r}_{10}\; kpc \cr
 M &= 10^{10} {\bar M}_{10} \; M_\odot
\label{rescaled2} } 
\end{equation}
Eq. (\ref{rotvel}) may now be written in a rescaled form as \be 
{\bar v}_{100}^2 ={1\over 2} {{\bar M}_{10} \over {\bar r}_{10}}- 
3 \times 10^{-5} {\bar \Lambda}_{52}\; {\bar r}_{10}^2  
\label{rescv}\ee In order to calculate the effects of the 
cosmological constant on the dynamically obtained masses of 
galaxies (including their halos) it is convenient to calculate the 
ratio 
  
\be p \equiv {{M ({\bar \Lambda}_{52} = 1) -  M({\bar 
\Lambda}_{52} = 0)}\over {M({\bar \Lambda}_{52} = 0)}}= {{3  
\times 10^{-5} {\bar r}_{10}^2} \over {{\bar v}_{100}^2}}  
\label{massrat}  \ee  where $M(\Lambda_{52})$ is the dynamically 
calculated mass obtained from equation (\ref{rescv}). 
 In Table 1 we show a calculation of the mass ratio $p$ for 22 
galaxies of different sizes and rotation 
velocities\cite{Sanders96}. The corresponding plot of $p(r)$ is 
shown in Fig. 1. 
\begin{figure}[ht]
\vskip.2in \mbox{\epsfxsize=10cm \epsffile{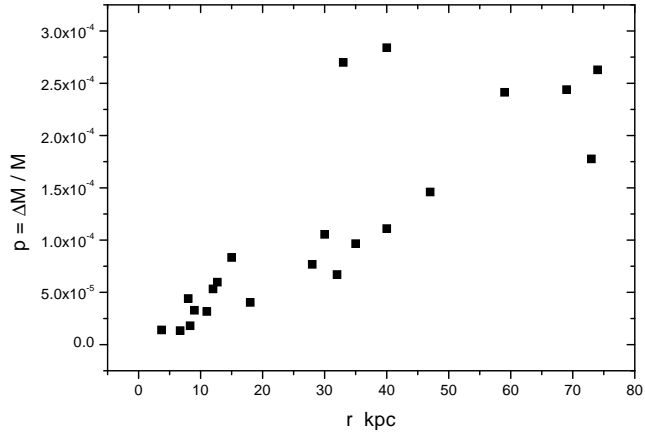}} 
\caption{Relative increase $p$ of dynamically calculated mass of 
galaxies due to repusive effects vacuum energy vs galaxy radious 
$r$ measured in kpc.} \label{fig1} 
\end{figure}

It is clear that even for large galaxies where the role of the 
repulsive force induced by vacuum energy is maximized, the 
increase of the mass needed to compensate vacuum energy 
antigravity is negligible. In order for these effects to be 
significant the cosmological constant would have to be larger that 
the value required for flatness by a factor of at least $10^3$. 

\begin{table}[ht]
\caption[Table 1]
{Relative increase $p$ of dynamically calculated mass of 
 galaxies due to repulsive effects of vacuum energy.}
\begin{tabular*}{\textwidth}{@{\extracolsep{\fill}}lccc}
\sphline \it {\bf Galaxy}&\it $ r_{HI}$  kpc &\it $v_{rot}$ km/sec 
&\it  $p$  \cr \sphline UGC2885&73&300&$17.8\times 10^{-5}$\\ 
 NGC5533&74&250&$26.3\times 10^{-5}$\\
 NGC6674&69&242&$24.3\times 10^{-5}$\\
 NGC5907&32&214&$6.7\times 10^{-5}$\\
 NGC2998&47&213&$14.6\times 10^{-5}$\\
 NGC801&59&208&$24.1\times 10^{-5}$\\
 NGC5371&40&208&$11.1\times 10^{-5}$\\
 NGC5033&35&195&$9.7\times 10^{-5}$\\
 NGC3521&28&175&$7.7\times 10^{-5}$\\
 NGC2683&18&155&$4.0\times 10^{-5}$\\
 NGC6946&30&160&$10.5\times 10^{-5}$\\
 UGC128&40&130&$28.4\times 10^{-5}$\\
 NGC1003&33&110&$27.0\times 10^{-5}$\\
 NGC247&11&107&$3.1\times 10^{-5}$\\
 M33&8.3&107&$1.8\times 10^{-5}$\\
 NGC7793&6.7&100&$1.3\times 10^{-5}$\\
 NGC300&12.7&90&$6.0\times 10^{-5}$\\
 NGC5585&12&90&$5.3\times 10^{-5}$\\
 NGC2915&15&90&$8.3\times 10^{-5}$\\
 NGC55&9&86&$3.2\times 10^{-5}$\\
 IC2574&8&66&$4.4\times 10^{-5}$\\
 DDO168&3.7&54&$1.4\times 10^{-5}$\\  \sphline 
\end{tabular*}

\end{table}
\inxx{captions,table} Such value would be inconsistent with 
several cosmological observations. Therefore, even though the 
effects of vacuum energy on galactic dynamics are much more 
important compared to the corresponding effects on solar system 
dynamics it is clear that we must consider systems on even larger 
scales where the mean density is smaller in order to obtain any 
nontrivial effects on the dynamics. 

The Local Group of galaxies is a particularly useful system for 
studying mass dynamics on large scales because it is close enough 
to be measured and modeled in detail yet it is large enough (and 
poor enough) to probe the effects of vacuum energy on the 
dynamics. The dominant members of the group are the Milky Way and 
the Andromeda Nebula M31. Their separation is \be r_0 \equiv r 
(t=t_0) \simeq 800 kpc  \label{galsep}\ee and the rate of change 
of their separation is  \be {{dr}\over dt}(t=t_0) \simeq -123 km 
s^{-1}  \label{galvel}\ee A widely used assumption is that the 
motion of approach of M31 and Milky Way is due to the mutual 
gravitational attraction of the masses of the two galaxies. 
Adopting the simplest model of the Local Group as an isolated two 
body system, the Milky Way and M31 have negligible relative 
angular momentum and their initial rate of change of separation is 
zero in comoving coordinates. The equation of motion for the 
separation $r(t)$ of the centers of the two galaxies in the 
presence of a nonzero cosmological constant is: 
\be
{{d^2 r}\over {dt^2}}= -{{G M}\over r^2} + {{\Lambda c^2}\over 3} 
r  \label{galacc}\ee  where M is the sum of the masses of the two 
galaxies. A similar equation (with $\Lambda = 0$) was used in Ref 
\cite{p93} to obtain an approximation of the mass to light ratio 
of the galaxies of the Local Group. Numerical studies\cite{pmhj89} 
have shown that this approximation is reasonable and leads to a 
relatively small overestimation (about 25\%) of the galactic 
masses. This correction is due to the effects of the other dwarf 
members of the Local Group that are neglected in the isolated two 
body approximation. Here we are not interested in the precise 
evaluation of the masses of the galaxies but on the effects of the 
cosmological constant on the evaluation of these masses. Therefore 
we will use the `isolated two body approximation' of the Local 
Group (eq. (\ref{galacc})) and focus on the dependence of the 
calculated value of the mass $M$ as a function of $\Lambda$ in the 
range of cosmologicaly interesting values of $\Lambda$.  Our goal 
is to find the total mass $M$ of the Local Group galaxies, using 
eq. (\ref{galacc}) supplied with the following conditions: \bea  
r(t=t_0) &=& 800\; kpc \label{condition1}\\ 
 {{dr}\over {dt}}(t=t_0) &=& -123 \; km/sec \label{condition2} \\
{{dr}\over {dt}}(t=0) &=& 0 \label{condition3} \eea  where $t_0 =  
15Gyr$. Upon integrating and rescaling eq. (\ref{galacc}) we 
obtain \be ({{d{\bar r}_{100}}\over {d{\bar t}_{15}}})^2 = {\bar 
M} ({1\over {{\bar r}_{100}}} - {1\over 8}) + {\bar \Lambda} 
({\bar r}_{100}^2 -64) + 420 \equiv f({\bar 
r}_{100})\label{resclg} \ee where we have used condition 
(\ref{condition2}) and the rescaled quantities defined by \bea  r 
&=& 100 \; {\bar r}_{100} \;  kpc \\ t &=& 1.5 \times 10^{10}\; 
{\bar t}_{15} \; yrs \\ M &=& 4\times 10^{8}\; {\bar M}\;  
M_\odot\\ \Lambda &=& 1.3 \; {\bar \Lambda}_{52} 
\label{rescalings} \eea Using now conditions (\ref{condition1}) 
and (\ref{condition3}) we obtain the equation that can be solved 
to evaluate the galactic masses for various $\Lambda$ \be 1 ={\bar 
t}_{15} (t=t_0) = - \int_{{\bar r}_{100}(t=0)}^{{\bar 
r}_{100}(t=t_0)} {{dr}\over \sqrt{f({\bar r}_{100})}} 
\label{intlg} \ee The lower limit of the integral (\ref{intlg}) is 
obtained by solving condition (\ref{condition3}) for $r$ (using 
eq. (\ref{resclg}) while the upper limit is given by eq. 
(\ref{condition1}) in its rescaled form. This equation can be 
solved numerically for $M$ to calculate the galactic total mass 
$M$ for various values of the cosmological constant $\Lambda$.  
The resulting dependence of $M$ on $\Lambda$ is shown in Fig. 2 
(continuous line).

\begin{figure}[ht]
\vskip.2in \mbox{\epsfxsize=10cm \epsffile{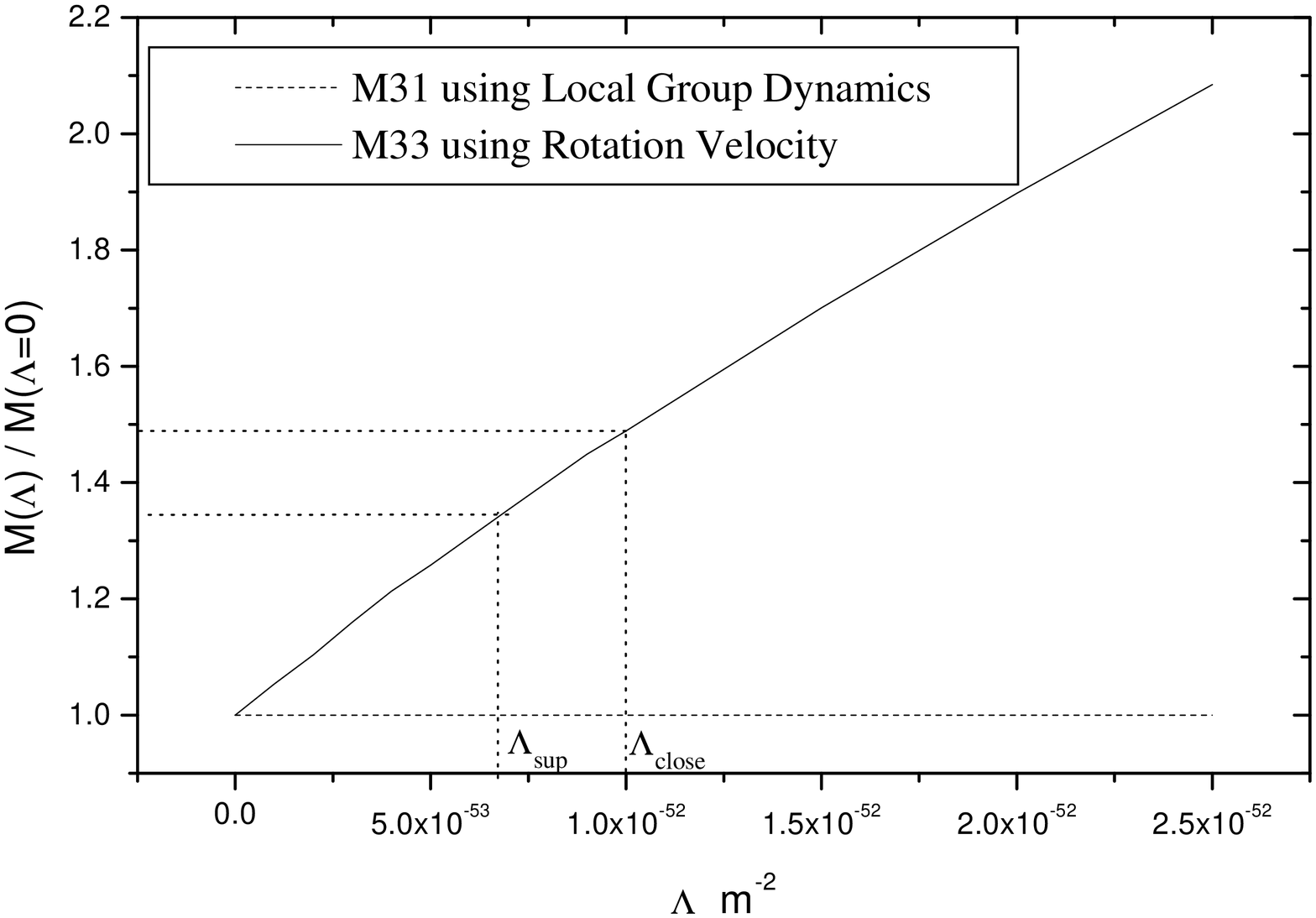}} \caption{The 
relative galactic total mass ${{M(\Lambda)}\over {M(\Lambda=0)}}$ 
calculated for various values of the cosmological constant 
$\Lambda$ using Local Group (continous line) and galactic scale 
(M33, dashed line) velocity data. The dotted lines correspond to 
the $\Lambda$ value implied by the SnIa data and to $\Lambda_{cl}$ 
for which the vacuum energy alone closes the universe.} 
\label{fig2} 
\end{figure}


Clearly, for a value of $\Lambda$ consistent with the recent SnIa 
observations ($\Lambda\simeq 0.7\times 10^{-52}m^{-2}$) the 
calculated galactic masses using Local Group dynamics are 35\% 
larger than the corresponding masses calculated with $\Lambda = 
0$. On Fig. 2 we also plot (dashed line) the dependence of the 
${{M(\Lambda)}\over {M(\Lambda = 0)}} $ on $\Lambda$ calculated 
using galactic dynamics (rotation velocity) of the galaxy M33. 
Clearly a galactic system in contrast to the Local Group is too 
small and dense to be a sensitive detector of the cosmological 
constant.

\section{Discussion}
We conclude that the Local Group of galaxies is a system that is 
large enough and with low enough matter density to be a sensitive 
probe of the gravitational effects of a cosmological constant with 
value consistent with cosmological expectations and recent SnIa 
obsrevations. Smaller and denser systems do not have this 
property. On the other hand the gravitational effects of $\Lambda$ 
should be even more pronounced on larger cosmological systems. 

The fact that the dynamical effects of negative pressure are 
manifested on scales of about $1Mpc$ and larger opens up 
interesting new windows for the determination of the density and 
equation of state of dark energy. 
%
For example the gravitational effects discussed here can be used 
as an independent detection method of the cosmological constant, 
if the galactic masses of systems like the Local Group are 
measured independently using non-dynamical methods (eg 
gravitational lensing).  In that case $\rho_\Lambda$ could be 
obtained using plots like the one shown in Fig. 2. Also the 
unexpected quietness and linearity of the local Hubble flow in the 
very clumpy local universe ($1-10Mpc$) can be attributed to the 
negative pressure\cite{bct00} of the dark energy and used to 
constrain $\rho_Q$ and $w$. 

\section{Acknowledgements}
We would like to thank M. Plionis for useful conversation.

\end{document}